\documentclass[fleqn,12pt,twoside]{article}
\usepackage{espcrc1}

\usepackage{graphicx}

\def\aJLone<#1,#2>{#1}
\def\aJLtwo<#1,#2,#3>{#2}
\def\aJLyear<#1,#2,#3,#4>{#3}
\def\aJLpage<#1,#2,#3,#4>{#4}
\def\JL#1{\aJLone<#1>\ {\bf \aJLtwo<#1>} (\aJLyear<#1>) \aJLpage<#1>}
\def\aJpage<#1,#2,#3>{#3}
\def\andvol#1{{\bf \aJLone<#1>} (\aJLtwo<#1>) \aJpage<#1>}
\def\PTP#1{Prog.\ Theor.\ Phys.\ \andvol{#1}}

\def\PR#1{Phys.\ Rev.\ \andvol{#1}}
\def\PRL#1{Phys.\ Rev.\ Lett.\ \andvol{#1}}
\def\PL#1{Phys.\ Lett.\ \andvol{#1}}
\def\NP#1{Nucl.\ Phys.\ \andvol{#1}}

\newcommand{\bra}{\langle}
\newcommand{\ket}{\rangle}
\newcommand{\braket}[1]{\bra #1 \ket}

\newcommand{\sGs}{\braket{\bar{s}g_s \sigma\cdot G s}}

\newcommand{\MB}{M}

\newcommand{\TP}{\Theta^+}
\newcommand{\sth}{\sqrt{s_{\rm th}}}

\newcommand{\Tate}{\rule{0cm}{1.1em}}

\title{QCD Sum Rules of Pentaquarks}

\author{Makoto Oka\address[Titech]{Department of Physics, H-27, Tokyo Institute of Technology \\ 
        Meguro, 152-8551, Japan}%
        \thanks{email: oka@th.phys.titech.ac.jp}
		}
       
\begin{document}

\maketitle

\begin{abstract}
QCD sum rule is applied to the pentaquark spectroscopy.  It is concluded that
no positive parity state is seen in low energy region, while there may exist negative parity states
at around 1.5 GeV.  
Choice of interpolating local operators and relation to the lattice calculations
are discussed.
\end{abstract}

\section{Introduction}

The newly discovered pentaquark $\Theta^+$\cite{Nakano,Hicks} 
and its siblings are quite mysterious from the QCD
viewpoint.
Many questions are raised including
\begin{itemize}
\item Why are they so light? The masses $M(\Theta^+) = 1540$ MeV and $M(\Xi^{- -}) = 1862$ MeV are
unexpectedly small as states with five constituent quarks.
\item Why are they so narrow?  The width $\Gamma(\Theta^+) < 10$ MeV is unusually narrow.
This may be the biggest problem in understanding the structure of $\Theta^+$.
\item What are the spin, parity, SU(3) irrep., \ldots? 
\item How are they produced?  This is a question raised from the beginning, but 
recent reports of ``non-observation'' of pentaquarks
make this question more important.  In which process do we have to see the state?
\end{itemize}

The spin and parity seem to be  crucial to determine the  structure of the pentaquarks.
In the naive quark models, the ground state in which all the quarks occupy the lowest energy mode
should have negative parity.  We expect both the $1/2$ or $3/2$ states, whose splitting is given by
the spin-spin interaction between quarks.\cite{Carlson,Shinozaki}
	
On the other hand, various model calculations predict a positive parity state.
In the chiral quark model,\cite{Diakonov} whose prediction motivated the experimental search at the SPring-8,
the expected state has $1/2^+$.  Another popular type of models are diquark models, in which 
a $ud$ ($S=0$, $I=0$) diquark plays a crucial role.\cite{JW,Enyo} 
The ground states in those models are predicted to
be a positive parity state with spin 1/2 or 3/2.

It is thus very important to determine the parity, but experimentally its determination 
turns out to be non-trivial.\cite{Thomas}
So while our experimental colleagues are working hard to determine
the spin and parity of the pentaquarks, theorists should make predictions directly from QCD.
This is the subject of this talk.

The main part of this talk is based on the research carried out by the Tokyo Tech.\ group 
including J.~Sugiyama, T.~Doi,\cite{SDO} N.~Ishii, H.~Iida, F.~Okiharu, and H.~Suganuma.\cite{Ishii}

\section{Two-point Correlator in QCD}	
	
A tool to attack the pentaquark question in QCD is two-point correlator (TPC).
TPC defined by
\begin{eqnarray}
\Pi (p) &=& \int d^4x e^{ip\cdot x} \langle 0| T(J(x) \bar J(0)) |0\rangle 
\end{eqnarray}
contains all information of the spectrum of the channel specified by the employed
interpolating field local operator, $J(x)$.

In the lattice QCD, TPC with the imaginary time variable $\tau$ is considered,
By integrating over 3-D coordinate $\vec x$ and taking the large $\tau$ limit, 
it will give the mass of the ground state.
\begin{eqnarray}
\Pi (\vec p=0; \tau) &=& \int d^3x\, \langle 0| T(J(\vec x,\tau) \bar J(0,0)) |0\rangle \nonumber\\
&=& \sum_i |\lambda_i|^2\,  e^{-m_i \tau} \quad \stackrel{\hbox{large $\tau$}}{\longrightarrow} 
\quad |\lambda_0|^2 \, e^{-m_0\tau}
\end{eqnarray}

The QCD sum rule is a relation obtained by calculating TPC in two ways.
In one hand, it is calculated in the deep Euclidean region 
$-p^2 \equiv p_E^2 \to \infty$.
This is the region where the perturbation theory is valid because 
the QCD coupling constant $\alpha_s(p_E^2)$ is suppressed according to the
asymptotic freedom of QCD.
In fact, QCD allows a systematic expansion called the operator product expansion (OPE)
in this regime and 
nonperturbative effects are taken into account as vacuum condensates of local operators.
\begin{eqnarray}
\Pi(p_E^2) &=& \sum_n C_n(p_E^2) \langle 0| O_n |0 \rangle
\end{eqnarray}
where $O_n$ represents a local operator of dimension $n$.  For higher $n$,
$C_n$ is suppressed as the $n$-th power of $1/p_E^2$ at large $p_E^2$, which  leads to
convergence of OPE.

The other side of the sum rule is TPC calculated at the physical region, where it is
represented by a spectral function $\rho (s)$  at $s = p^2 = m^2$:
\begin{eqnarray}
 \rho(s) &=& \frac{1}{\pi} {\rm Im} \Pi (s)= \sum_i |\lambda_i|^2 \delta(s-m_i^2) 
\end{eqnarray}
where $i$ labels hadrons (poles) and scattering states (cut).
Incidentally a common assumption in the QCD sum rule is to parametrize 
the spectral function by a sharp resonance and continuum at $s>s_0$,
\begin{eqnarray}
\rho(s) &\sim& |\lambda|^2 \delta(s-m^2) + \theta (s-s_0) \, \rho(s)_{\rm OPE} 
\qquad \hbox{$s_0$: threshold}
\label{eq:rho}
\end{eqnarray}

Then the two expressions of TPC at different kinematical regimes are related by using
analyticity of $\Pi(p^2)$ in complex $s=p^2$ variable.
The analytic continuation is achieved by a dispersion relation,
\begin{eqnarray}
\Pi (p_E^2) &=& \frac{1}{\pi} \int_0^{\infty} ds \, \frac{{\rm Im}\, \Pi (s)}{s+p_E^2} .
\end{eqnarray}

In order to improve sum rule performance, it is common to apply the Borel transform defined by
\begin{eqnarray}
\Pi(p^2= -p_E^2)  \to {\cal B}_{M^2}\Pi \equiv \tilde\Pi(M^2) = 
\lim_{p_E^2, n \to \infty,  M^2 \equiv p_E^2/n = {\rm finite}}
{(p_E^2)^{n+1}\over n!} \left(-{d\over dp_E^2}\right)^n \Pi(p_E^2) ,
\end{eqnarray}
where $M^2$ is a mass scale introduced in the transform and is called Borel mass.
An example of the Borel transform is
\begin{eqnarray}
{\cal B}_{M^2} \int_0^{s_0} {\hbox{Im} \Pi(s)\over s+p_E^2} ds =
\int_0^{s_0} e^{-s/M^2} \hbox{Im} \Pi(s)ds ,
\end{eqnarray}
which shows its role clearly, that is, it enhances the pole region by the
weight function $e^{-s/M^2}$.
It also accelerates the convergence of OPE by an extra numerical factor proportional to $1/n!$.

\subsection{Parity Projection}

Our main aim is to determine the parity of pentaquarks, for which we need parity projection.
Parity projection for $J = 1/2$ baryons was formulated by Jido et al.\cite{JKO}
There the authors employ the forward-time TPC,
\begin{eqnarray}
\Pi (p) &=& \int d^4x e^{ip\cdot x} \theta(x^0) \langle 0| J_B(x) \bar J_B(0) |0\rangle ,
\end{eqnarray}
then its imaginary part at $\vec p=0$ can be projected into positive parity and negative
parity parts as
\begin{eqnarray}
{\rm Im} \Pi (p_0, \vec p=0) &=& |\lambda_+|^2 \frac{\gamma_0+1}{2} \delta(p_0-m_+) +
|\lambda_-|^2 \frac{\gamma_0-1}{2} \delta(p_0-m_-)  + \hbox{ (continuum)}\nonumber\\
&=& \gamma_0\, A(p_0) + B (p_0) .
\end{eqnarray}
OPE of the correlator shows that $A(p_0)$ contains chiral even parts,
such as the gluon condensate, while 
$B(p_0)$ corresponds to the chiral odd parts, such as the quark condensate, 
$\langle \bar q q\rangle$. 

Now one sees that the combinations, $A+B$ and $A-B$, represent spectral functions 
of the positive and negative parity baryons, respectively.
By constructing the QCD sum rule for $A\pm B$,
one can extract masses and coupling strengths of the positive and negative
parity baryons.
It is concluded that splitting of positive and negative parity baryons is determined by
chiral odd terms in OPE, and thus chiral symmetry breaking of the QCD vacuum.
The final sum rules for the positive and negative parity baryons are given by
\begin{eqnarray}
&&\int_0^{s_+} \,[A_{\rm OPE}(p_0) + B_{\rm OPE}(p_0) ] \exp\left[ - \frac{p_0^2}{M^2}\right] \, dp_0
= |\lambda_+|^2 \exp\left[-\frac{m_+^2}{M^2} \right]\\
&&\int_0^{s_-} \,[A_{\rm OPE}(p_0) - B_{\rm OPE}(p_0) ] \exp\left[ - \frac{p_0^2}{M^2}\right] \, dp_0
= |\lambda_-|^2 \exp\left[-\frac{m_-^2}{M^2} \right]
\end{eqnarray}

\subsection{Choice of Interpolating Local Operator}

In principle, any local operator with appropriate quantum numbers can be
employed for sum rules, or lattice QCD calculations. It is, however, not
the case in practice.  Because one has continuum contribution from hadronic
background, it is important to enhance the pole term as much as possible 
by choosing the most appropriate operator.

This situation is well demonstrated in the case of the sum rule and 
the lattice calculation for the nucleon.
It is known that there are two independent local operators (without derivatives)
for the nucleon, 
\begin{eqnarray}
B_1(x) &=& \epsilon_{abc} [u_a^T(x)C\gamma_5 d_b(x)]\, u_c(x)\nonumber\\
B_2(x) &=& \epsilon_{abc} [u_a^T(x)C d_b(x)]\, \gamma_5 u_c(x) ,
\end{eqnarray}
where $C$ denotes the gamma matrix for charge conjugation and $a, b, \ldots$ are color indices.
Then a general 3-quark operator can be written as
\begin{equation}
 J(x) = B_2(x)+ t B_1(x) 
\end{equation}
$B_1$ and $B_2$ have different chiral properties, which can be seen in some particular
limits:
For $t=-1$, $J$ belongs to the $(3,\bar 3) + (\bar 3,3)$ representation of SU(3)$_L\times$SU(3)$_R$, which is called the Ioffe current.
For $t=+1$, it belongs to $(8,1)+ (1,8)$, and for
$t=0$, it is purely $B_2$, which vanishes in the NR limit.
It is practically important in lattice QCD and sum rules to determine 
which is the best operator for the positive parity and negative
parity nucleons.

Such a study was performed for lattice QCD by Sasaki et al.\cite{Sasaki1}
by using the domain wall fermion formalism.
They demonstrated that the $B_1$ operator does not couple to the nucleon $N(940)$ 
and gives its excited state $N^*(1440)$ fairly well.
An analysis for the sum rule was performed by Jido.\cite{Jido}  
He showed that the positive parity baryon has strong coupling to the Ioffe current, 
while the negative parity strength is most strong for a positive $t$, i.e., $t\sim 0.8, 1.2$.

\section{Sum Rule for Pentaquark}
A local operator for $\Theta^+$, employed by Ref.\cite{SDO}, is
\begin{eqnarray}
\Theta^+ (x) &\equiv& 
\epsilon_{abc} \epsilon_{ade} \epsilon_{bfg} [u^T_d(x) C\gamma^5 d_e(x)][u^T_f(x) C d_g(x)]
C \bar s^T_c(x) .
\end{eqnarray}
This operator is chosen because its overlap with $NK$ states is suppressed,
which can be seen from the Fiertz rearrangement,\cite{Ishii}
\begin{eqnarray}
\Theta^+(x) 
&=&
  \epsilon_{ade}
  \left( u^T_d C\gamma_5 d_e \right)
  \times
  \frac1{4}
  \left\{\Tate\right.
  - \gamma_5 d_a \left( \bar{s}_c\gamma_5 u_c \right)
  + \gamma_5 u_a \left( \bar{s}_c\gamma_5 d_c \right)
  - d_a \left( \bar{s}_c u_c \right)
  + u_c \left( \bar{s}_c d_c \right)
  \nonumber\\
  &&
  + \gamma_{\mu} d_a \left( \bar{s}_c\gamma_{\mu} u_c \right)
  - \gamma_{\mu} u_a \left( \bar{s}_c\gamma_{\mu} d_c \right)
  + \gamma_5\gamma_{\mu} d_a \left( \bar{s}_c\gamma_5\gamma_{\mu} u_c \right)
  - \gamma_5\gamma_{\mu} u_a \left( \bar{s}_c\gamma_5\gamma_{\mu} d_c \right)
  \nonumber\\
  &&
  + \frac1{2} \sigma_{\mu\nu} d_a \left( \bar{s}_c\sigma_{\mu\nu} u_c \right)
  - \frac1{2} \sigma_{\mu\nu} u_a \left( \bar{s}_c\sigma_{\mu\nu} d_c \right)
  \left.\Tate\right\}.
\end{eqnarray}

The residue of the pole, $|\lambda|^2$,  in Eq.~(\ref{eq:rho}) represents
the strength with which the interpolating operator couples to the physical state, 
and it should be positive if the pole is real.
We use this condition to determine the parity of the pentaquark.  
In Fig.~1, we plot the OPE side (as a function of $\MB$) corresponding to
$|\lambda_{\pm}|^2 \exp (-m^2/\MB^2)$.
We find that the dimension-5 condensate, $\sGs$, gives a large negative contribution to 
$|\lambda_{+}|^2$, which makes $|\lambda_{+}|^2$ to be nearly zero or even slightly negative.  
This suggests that the pole in the positive-parity spectral function is spurious.
In contrast, the large $\sGs$ contribution makes $|\lambda_{-}|^2$ positive.
We thus conclude that
the obtained negative-parity state is a real state.

\begin{figure}[hbtp]
\begin{center}
\includegraphics[width=8cm]{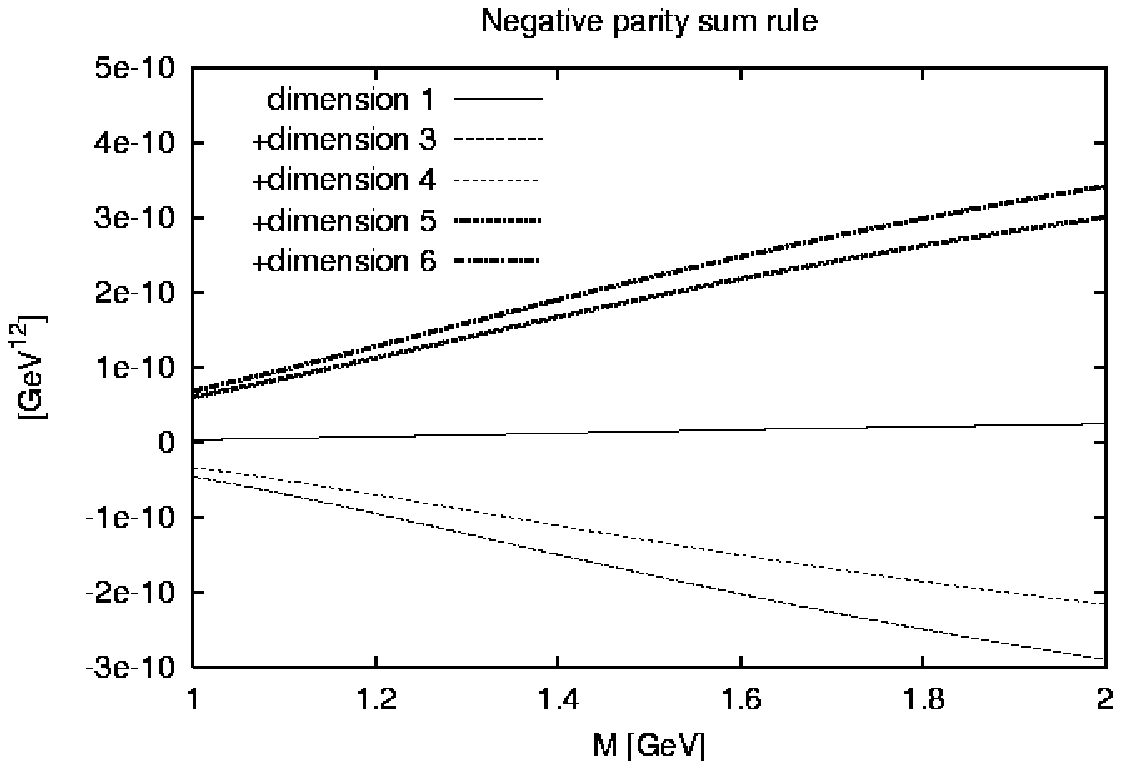}
\includegraphics[width=8cm]{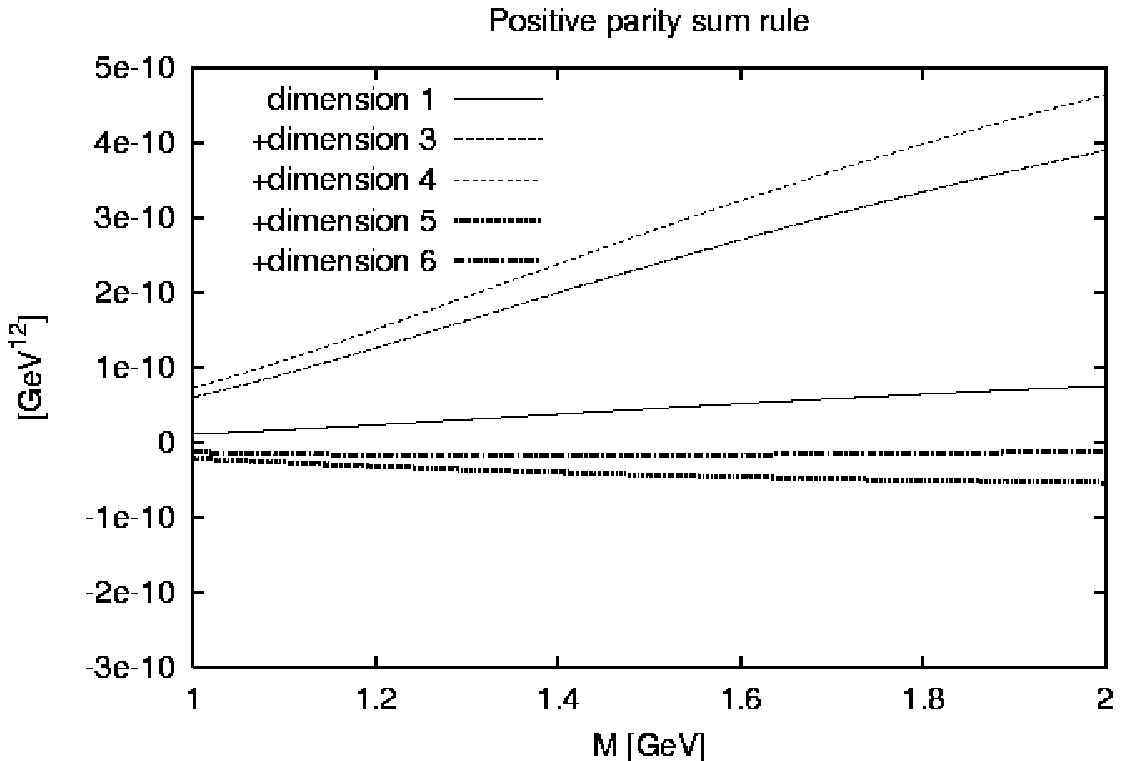} 
\caption{
Contributions from the terms of each dimension added up successively
for the negative-parity and positive-parity sum rules with $\sth=1.8$ GeV.%
}
\end{center}
\label{fig:R}
\end{figure}
The mass of the negative parity $\TP$ is estimated using the sum rule, and we find
that the $\MB$ dependence is rather weak. Therefore the reliability of the sum rule 
seems good.
The mass is, however, sensitive to choices of the continuum threshold $s_{th}$.
Nevertheless, we confirm that the result is consistent with the observed value.

Sum rules for other pentaquark baryons, such as $\Xi^{--}(I=3/2)$,
have been constructed. 
The results are similar to $\Theta^+$, that is, 
only negative parity correlator has significant strength below 2 GeV.
We also find that SU(3) breaking effects seem small in the negative-parity pentaquark states.
See Ref.\cite{Sugiyama} for details.

\section{Discussion}

How do we interpret the results?
Are they consistent with what our experimental colleagues have observed?
Are they really pentaquark states or something else?

What we see in the QCD sum rule may or may not be a sharp resonance state, because the sum rule 
cannot tell the width of the state.  (Coupling strength to decay channels may be calculated, which 
is related to the width indirectly.)
It is important that the interpolating local operator has small overlap with the
decay channel so that the signal to background ratio is enhanced.
It has been shown by Lee\cite{SHLee} in this workshop that contribution of
the $NK$ states in the non $NK$-type correlator is sufficiently small.
He indeed demonstrated that using the soft-meson technique, the $\Theta$-vacuum
matrix element of the local operator can be rewritten as a $N$-vacuum matrix 
element of a 5-quark nucleon operator.  
\begin{eqnarray}
\langle 0| J_{\Theta}|NK \rangle &=& - \frac{i}{f_{\pi}} 
\langle 0| [Q_5^K, J_{\Theta}] |N \rangle 
= - \frac{i}{f_{\pi}} \langle 0| J_{N;5q}|N \rangle
\end{eqnarray}
Constructing a (new) nucleon sum rule using the 5-quark operator in the right hand side,
Lee found that it is consistent with positive parity nucleon and that 
the $NK$ continuum contribution is less than 5\% 
in the $\Theta^+$ sum rule.

Similarly, choices of the operator may be critically important in lattice QCD calculations.
In the workshop, several lattice calculations of the mass of $\Theta^+$ 
with various local operators have been 
presented.\cite{Sasaki2,FLee,Chiu,Ishii,Takahashi}
All but one agree that signal of $1/2^+$ state is found only at very high mass, i.e., 
above 2 GeV.
They disagree, however, on the existence of a $1/2^-$ state.  
Lee\cite{FLee} presented their results of an extensive 
study by using the operator that is a product of $N$ and $K$.  
They claim that no signal for a compact 5-quark state is found.
It is naturally expected that the $NK$ 
operator does couple to the continuum state strongly and therefore is not appropriate to 
look for a signal of a compact resonance state.  
In contrast, Sasaki\cite{Sasaki2} carried out 
a pioneering study employing the
non $NK$ operator and claimed that TPC shows a double-plateau structure, one of which 
corresponds to the resonance above the other one, the $NK$ threshold.

Ishii et al.\cite{Ishii} studied $\Theta^+$ using an anisotropic lattice in order to enhance 
precision of the mass determination.  They employ quenched approximation with
the non $NK$ interpolating operator.  It is found a smeared source is useful in
extracting the lowest energy state effectively.
They have also introduced a new technique called hybrid boundary method and 
demonstrated that existence of a $1/2^-$ state is unlikely near the  $NK$ threshold.
Details of their results are given in ref.\cite{Ishii}.

\subsection{PP Strikes Back?}

Several authors have suggested existence of low-lying positive parity state in QCD.
In lattice QCD, Chiu and Hsieh\cite{Chiu} carried out quenched calculation with the domain-wall quark and concluded that a positive-parity state appears as the ground state.  
But their calculation contradicts with
the others and it is fair to say that the conclusion is pending right now.

QCD sum rule calculations also predicted the change of the order of the parity. 
Kim et al.\cite{Kim} constructed a sum rule for $\bar c$ pentaquark baryon
and found that a positive
parity state comes lower than negative parity states.  They found also that 
parity inversion has occurred due to new terms with the charm quark mass and therefore
their result is consistent with negative-parity $\Theta^+$.

Another study by Kondo et al.\cite{KMN} claimed that the conventional sum rule is contaminated 
by ``baryon-meson reducible'' diagrams and removal of such diagrams may reverse the conclusion on
the parity assignment.
They define a so-called two-hadron irreducible TPC as
\begin{eqnarray}
\langle T(J_{\Theta}(x) \bar J_{\Theta}(0) \rangle^{2HI} &=& \langle T(J_{\Theta}(x) \bar J_{\Theta}(0) \rangle
- \sum_{ij} \langle T(J^i_{B}(x) \bar J^j_{B}(0) \rangle \langle T(J^i_{M}(x) \bar J^j_{M}(0) \rangle ,
\end{eqnarray}
where the second term represents contribution of two hadrons propagating without interacting each other.
Then as two-hadron reducible terms they assign the diagrams in OPE that can split into two color-singlets with no interaction between them.

In principle, it is important to suppress contribution from $NK$ (and other hadronic) 
scattering states as much as
possible to isolate a sharp resonance state on top of it.
It is, however, shown that the subtraction of $NK$ reducible part is not simply accomplished by
throwing away the diagrams which have no connection between two color singlet parts.
The following problems are pointed out.
\begin{enumerate}
\item The 3-quark and $q\bar q$ operators are connected at the vertex,
where the quark operators are normal-ordered so that
divergence from vertex corrections is subtracted. 
Namely, a renormalization is required to isolate noninteracting hadrons, 
\begin{eqnarray}
J_{\Theta}(x)  = J_N(x) J_K(x)  \times Z_{\rm ren} + \ldots
\end{eqnarray}
Therefore the subtraction of the two-hadron reducible part should also take care of
the renormalization factor, which cannot be done simply by eliminating some of
the perturbative QCD diagrams.
\item It is also pointed out that analytic continuation should bring full interactions
among quarks (and gluons) in QCD sum rule and their interactions are determined by the local vertices.
\item Another problem is that the 3-quark and $q\bar q$ TPCs are not independent in the
sum rule, because quark (gluon) condensates of one correlator and the other should be
correlated with each other. 
\item The validity of the OPE for the two-hadron reducible part was pointed out by Lee.\cite{SHLee}
\end{enumerate}
Thus defining ``non-interacting'' part is not trivial.  
It is therefore concluded that subtraction of
non interacting part should be done more carefully, even if it is possible, and the results of 
ref.\cite{KMN} seem not correct.  The same conclusion was reached by Lee.\cite{SHLee}

\section{Conclusion}

The conclusion from QCD as of today is simple.
\begin{enumerate}
\item QCD predicts no $J^{\pi} = 1/2^+$ ($F =10^*$)  pentaquark. Most results indicate
its mass to be 2 GeV or higher.
\item Some calculations predict negative parity pentaquark state, but it may well 
be buried in the $NK$ continuum.
It certainly requires confirmation.
\end{enumerate}

What are possible remedies for the discrepancy between the QCD predictions and most
other model calculations.
Are there strong pionic effects, which may not be taken into account properly in the sum rule nor 
the quenched QCD calculations? It is, however, noted that the Skyrmion model predicts no less
``pionic'' effects in the nucleon and $\Delta$.  Why, then, the QCD calculations, even the quenched 
approximation, do so well for the ordinary baryons?

Another possibility is that the interpolating local operator is completely wrong. Such a possibility
may include that this is a state with 7 quarks or more.  If the state is a $NK\pi$ bound state, for 
instance, a 5-quark lattice QCD calculation hardly reproduces it. 
It should be interesting to look for some non QCD possibility for the ``pentaquark'' state.


\begin{thebibliography}{99}

\bibitem{Nakano} T. Nakano et al., \PRL{91,2003,012002}.
\bibitem{Hicks} K.~Hicks, Summary talk of the Workshop in this Proceedings.
\bibitem{Carlson} C.E.~Carlson, C.D.~Carone, H.J.~Kwee and V.~Nazaryan, 
\PL{B573,2003,101}; 
Fl.~Stancu and D.O.~Riska, \PL{B573,2003,242};
Fl.~Stancu, \PL{B595,2004,269}; 
B.K.~Jennings and K.~Maltman, \PR{D68,2004,094020};
R.~Bijker et al., hep-ph/0310281.
\bibitem{Shinozaki} T.~Shinozaki, S.~Takeuchi and M.~Oka, hep-ph/0409103.
\bibitem{Diakonov} D.~Diakonov, V.~Petrov and M.~Polyakov, \JL{Z.~Phys.~A,359,1997,305}.
\bibitem{JW} R.L.~Jaffe and R.~Wilczek, \PRL{91,2003,55};  M.~Karliner and H.J.~Lipkin, \PL{B575,2003,249}.
\bibitem{Enyo} Y.~Kanada-En'yo, O.~Morimatsu and T.~Nishikawa, hep-ph/0404144.
\bibitem{Thomas} A.W.~Thomas, K.~Hicks and A.~Hosaka, \PTP{111,2004,291};
C.~Hanhart et al., \PL{B590,2004,39}.
\bibitem{SDO} J.~Sugiyama, T.~Doi and M.~Oka, \PL{B581,2004,167}.
\bibitem{Ishii} N. Ishii et al., hep-lat/0408030.

\bibitem{JKO} D.~Jido, N.~Kodama and M.~Oka, \PR{D54,1996,4532}.
\bibitem{Sasaki1} S.~Sasaki et al., \PR{D65,2002,074503}.
\bibitem{Jido} D.~Jido, to be published; 
D.~Jido and M.~Oka, hep-ph/9611322; 
M.~Oka, D.~Jido and A.~Hosaka, \NP{A629,1998,156c} (hep-ph/9702351).
\bibitem{Sugiyama} J.~Sugiyama, T.~Doi and M.~Oka, in this Proceedings.

\bibitem{SHLee} S.H. Lee, in this Proceedings and to be published. 

\bibitem{Sasaki2} S.~Sasaki, hep-lat/0310014.
\bibitem{FLee} N.~Mathur et al., hep-ph/0406196.
\bibitem{Chiu} T.-W.~Chiu and T.-H.~Hsieh, hep-ph/0403020.
\bibitem{Takahashi} T. Takahashi, in this Proceedings.

\bibitem{Kim} H.~Kim, S.H.~Lee, Y.~Oh, \PL{B595,2004,293}.
\bibitem{KMN} Y.~Kondo, O.~Morimatsu and T.~Nishikawa, hep-ph/0404285.
\end{thebibliography}
\end{document}